\documentclass[proof]{pasj00}
\draft

\usepackage[dvipdfmx]{xcolor}

\newcommand{\xmm}{{\it XMM-Newton}\ }

\newcommand{\chandra}{{\it Chandra}\ }

\newcommand{\suzaku}{{\it Suzaku}\ }

\newcommand{\astroh}{ASTRO-H\ }

\begin{document}
\SetRunningHead{Author(s) in page-head}{Running Head}
\Received{2014/11/10}
\Accepted{2014/12/13}

\title{An X-ray Spectroscopic Search for Dark Matter in the Perseus Cluster with \suzaku}

\author{
Takayuki \textsc{Tamura},
{Ryo \textsc{Iizuka}},
{Yoshitomo \textsc{Maeda}},
{Kazuhisa \textsc{Mitsuda}} \\
and
{Noriko \textsc{Y. Yamasaki}}
}
\affil{
Institute of Space and Astronautical Science,
Japan Aerospace Exploration Agency,\\
3-1-1 Yoshinodai, Chuo-ku, Sagamihara, Kanagawa 229-8510, Japan
}
\email{tamura.takayuki@jaxa.jp}


%


\maketitle

\begin{abstract}
We present the results from deep \suzaku observations of the central region of the Perseus cluster.
\citet{bulbul2014} reported the detection by \xmm instruments of an unidentified X-ray emission line 
at an energy around 3.5~keV in spectra for the Perseus and other clusters.
They argued for a possibility of the decay of sterile neutrino, a dark matter candidate.
We examine \suzaku X-ray Imaging Spectrometer (XIS) spectra of the Perseus cluster
for evidence in the 3.5~keV line and other possible dark matter features in the 2--6~keV energy band.
In order to search for and constrain a weak line feature with the XIS,
observations of the Crab Nebula are used to evaluate the system's effective area.
We found no line feature at the claimed position
with a systematic line flux upper limit at a half (1.5~eV in line equivalent width) of the claimed best-fit value by Bulbul et al.
We discuss this inconsistency in terms of instrumental calibration errors and modeling of continuum emission.
Future prospects for high-energy resolution spectroscopy with \astroh are presented.
\end{abstract}

\section{Introduction}
\label{intro}

Dark matter was first suggested in galaxy clusters by Zwicky in the 1930s 
and was deduced from galactic rotation curves by Rubin and others in the 1970s (see \cite{Sofue2001} for a review).
Since then,
its spatial distribution
has been measured by various independent methods
over scales from small galaxies to galaxy clusters and beyond.
In ground laboratories, in particle colliders, and in space,
many experiments have attempted direct dark matter detection or creation.
Despite these efforts, 
no robust detection has yet been made
and the nature of dark matter is still mysterious.
Independent considerations
consistently indicate that particles within the SU(3)$\times$ SU(2)$\times$ U(1) Standard Model of particle physics
cannot comprise the main body of dark matter,  
suggesting a new physics beyond the Standard Model (see e.g. \cite{Feng2010} for a recent review).

Because of their extremely low interactions, 
some dark matter candidates can be detected only in decays or annihilation 
from cosmic systems with high enough numbers of particles.
As one of method of indirect astronomical study, 
\citet{Abazajian2001} proposed a search 
for the X-ray decay line of a hypothetical sterile neutrino, 
a right-chiral counterpart of the active neutrinos.
A keV-mass sterile neutrino is a warm dark matter candidate.
This sterile neutrino X-ray has been examined in observations
of the Milky Way, other galaxies, and galaxy clusters (e.g. 
\cite{Kusenko2009},
\cite{Abazajian2012}, 
\cite{Boyarsky2012},
and \cite{Horiuchi2014}
) without any clear detections.

Bulbul et al. (2014; B14) found an unidentified line at around 3.5~keV
in the \xmm European Photon Imaging Camera (EPIC) MOS 
and pn detector's stacked X-ray spectra from many clusters.
They reported the highest flux from the central region of the Perseus cluster.
\citet{boy2014a} and \citet{boy2014b} also reported detections at the same energy 
in the Perseus cluster off-center, the M31 galaxy, and the Galactic center.
Both groups claimed that 
their observed line positions and fluxes are consistent with a sterile neutrino decay signal,
with cautions regarding systematic uncertainties in calibrations and contamination from atomic transitions.
These detections motivated a number of theoretical studies (e.g. \cite{Nakayama2014}).

The possible dark matter origin suggested by B14 is intriguing.
To examine the 3.5~keV line detection and origins,
we analyzed deep \suzaku XIS (X-ray Imaging Spectrometer) data of the Perseus cluster center.
This data set provides independent and possibly deeper and more accurate measurements.
After careful examination of systematic uncertainty from instrumental calibrations,
we found no line at the claimed position.
We have also constrained emission of any unidentified line features over the 2.0--6.3~keV energy range.

Throughout this paper,
we assume the following cosmological parameters, 
$H_0 = 70$ km s$^{-1}$Mpc$^{-1}$, 
$\Omega_\mathrm{m} = 0.3$, and $\Omega_\mathrm{\Lambda} = 0.7$.
At the cluster redshift of 0.0183, 
one arcminute corresponds to 22.2~kpc.
Unless otherwise stated, 
we use the 68\% ($1\sigma$) confidence level for errors
and the X-ray energy at the observed (hence redshifted) frame, 
rather than the object's rest-frame.

\section{Energy response calibration: Crab Nebula spectral fit}
\label{sect:crab}

\subsection{Motivation}
\label{sect:crab-mot}

In this paper, we examine deep \suzaku data for the 3.5~keV line emission suggested by Bulbul et al.'s (2014) \xmm analysis.
To find a weak line feature on top of bright continuum emission,
the energy response calibration is important, 
as are statistically robust data.
Fig.~\ref{ana-cal:arf} shows the effective area of {\it Suzaku}'s XIS and XRT (X-Ray Telescope) system.
There are strong and weak instrumental edges over the energy band.
Any uncertainties around these instrumental features introduce systematic errors in line detection.

The energy scale is also particularly important for identifying observed signals with known atomic lines.
Moreover, it is required to correct positions of strong instrumental features.
In \citet{tamura2011} and \citet{tamura2014}
we evaluated the energy scale in details for the analysis of the same Perseus data as the current study.
We found an energy scale uncertainty at a level of 0.1--0.2\% around the Fe-K energy.

The line spread function
is another factor which needs to be calibrated accurately.
This uncertainty could introduce systematic residuals around strong lines.
The Si-escape peaks of strong lines must also be taken into account.

\subsection{Reported Status}
\label{sect:cal-status}

\citet{koyama07} summarized ground and initial on-orbit XIS calibration.
\citet{ozawa09} and \citet{uchiyama09} reported long-term calibration for 
spaced-row charge injection (SCI) 'off' and 'on' mode observations, respectively.

The XRT part of the calibration was fully described in \citet{serlemitsos07}
who used Crab Nebula observations to evaluate the effective area and 
reported that the standard responses 
and a simple power-law model
can reproduce the Crab spectra observed at normal positions
with residuals less than 5--10\% over the 1.0--10~keV band.
They identified systematic errors around the Si-K edge at 1.84~keV.
Updated calibration is reported 
in a series of \suzaku memo by Maeda et al.
and summarized in ``The Suzaku Technical Description''
\footnote{These documents can be found at http://www.astro.isas.jaxa.jp/suzaku/doc/.}.
\citet{Ishida2011} and \citet{Tsujimoto2011}
reported the cross calibration 
of \suzaku and other X-ray observatories
using PKS~2155-304 and G21.5-0.9, 
respectively,
focusing on differences in the total flux and the spectral slopes.

\begin{figure}[hpt]
\begin{center}
\includegraphics[scale=.50]{fig-area.ps} 
\caption{
XIS effective area curve 
[sum of two frontside-illuminated charge-coupled devices (FI CCD), 
namely XIS-0 and -3 in the 1.5--4.0~keV band.
The backside-illuminated (BI) CCD has similar features.
Positions of Si-K and Au-M edges are indicated by lines.
(Plot is rotated due to a latex technical problem.)
}
\label{ana-cal:arf}
\end{center}
\end{figure}

\subsection{Crab Nebula Spectral Evaluation and Corrections}
\label{crab-obs}

To evaluate the effective area calibration,
we followed previous XRT reports 
given above and used \suzaku data for the Crab Nebula.
This nebula, 
a bright one having non-thermal emission without any line emission, 
is a calibration target 
for \suzaku instruments and was observed regularly in different observation modes and positions.

We used SCI 'on' mode data from 
sequences, 
103007010, 
103008010, 
104001070, 
105002010, 
105029010,
106012010,
106013010,
106014010,
106015010,
and 108011010.
The exposure time for each sequence is in the range 400--500~s.

Spectra are accumulated within $3'$ of the Crab Nebula's position from each charge-coupled devices (CCD) and each observation.
The front-side illuminated (FI) CCD data (XIS-0 and -3) are added together.
The instrumental background is below 0.1\% of the source in the 1--8~keV energy band.
The Crab exposures include so-called out-of-time events, 
which could have an energy response slightly different from normal events.
We confirmed that these special events have no significant impact on the fitting results in local bands.
Therefore no background or out-of-time events were subtracted.

The data are sorted by observation periods
into two groups, 
AO (announcement of opportunity) 3-4-5 and AO6-8.

For each group spectrum,
an XIS energy response function (redistribution matrix file or RMF) is prepared using
{\tt xisrmfgen} \citep{ishisaki07} for the corresponding observation period.
We used 
ae\_2FI\_xisnom6\_ao4\_20080910.arf 
and ae\_BI\_xisnom6\_ao4\_20080910.arf as XRT ancillary response function files (ARFs). 
These were released from the \suzaku project 
for a point source at the XIS nominal position.

\begin{figure}[hpt]
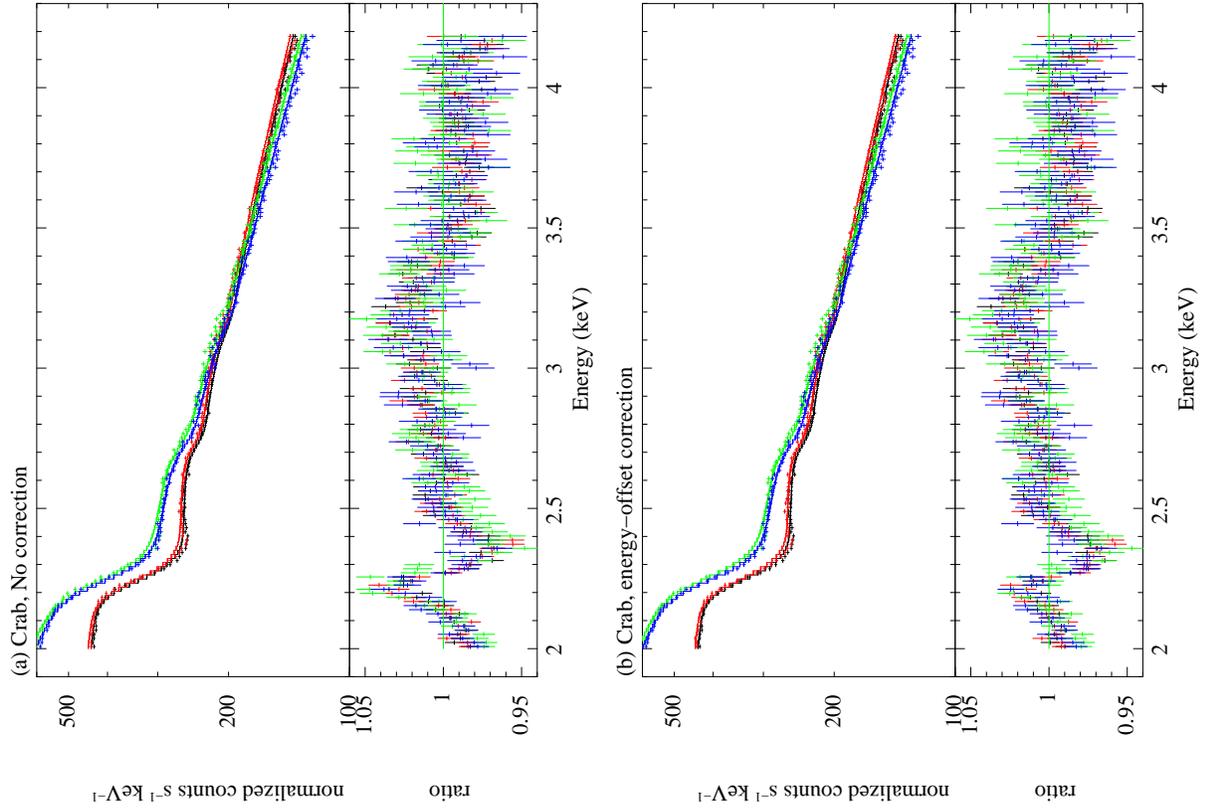

\begin{center}
\includegraphics[scale=.42]{fig-crabFit1.ps}
\includegraphics[scale=.42]{fig-crabFit2.ps}
\caption{
(a) 
XIS Crab spectra fitted with a single power-law modified by Galactic absorption ('standard response').
The black, red, blue, and green lines 
are the FI/AO3-4-5, FI/AO6-8,
BI/AO3-4-5 and BI/AO6-8 spectra, respectively.
The upper panel and lower panel show counts s$^{-1}$ keV$^{-1}$ and data-to-model ratio, respectively.
(b) Same as panel (a),
but with offset correction.
}
\label{crab:fit1}
\end{center}
\end{figure}

Since we are searching mainly for a line around 3.5~keV, 
we focus on the 2.0--4.4~keV band.
We fitted the spectra
with a power-law modified by a fixed Galactic absorption column density
of $3\times 10^{21}$ cm$^{-2}$.
As shown in Fig.~\ref{crab:fit1}, 
this 'standard response' model reproduces all the spectral channels
within $\pm$5\% accuracy, 
consistent with reports given above.

To compensate for 
possible energy-scale error, 
we use the 'gain' command in the XSPEC (version-12.8; \cite{Arnaud1996}) package 
to correct energy-scale offsets ('offset correction' in Table~\ref{crab:fit-para}).
As shown in Fig.~\ref{crab:fit1} and Table~\ref{crab:fit-para},
the fit improves significantly.
The obtained offset values are less than 12~eV (Table~\ref{crab:fit-para}).

Even after the offset tuning, 
two significant and systematic residuals can be found:  
a sharp feature around an Au M edge ($\sim$ 2.3~keV)
and a wider one around 2.8--3.5~keV.

\begin{figure}[hpt]
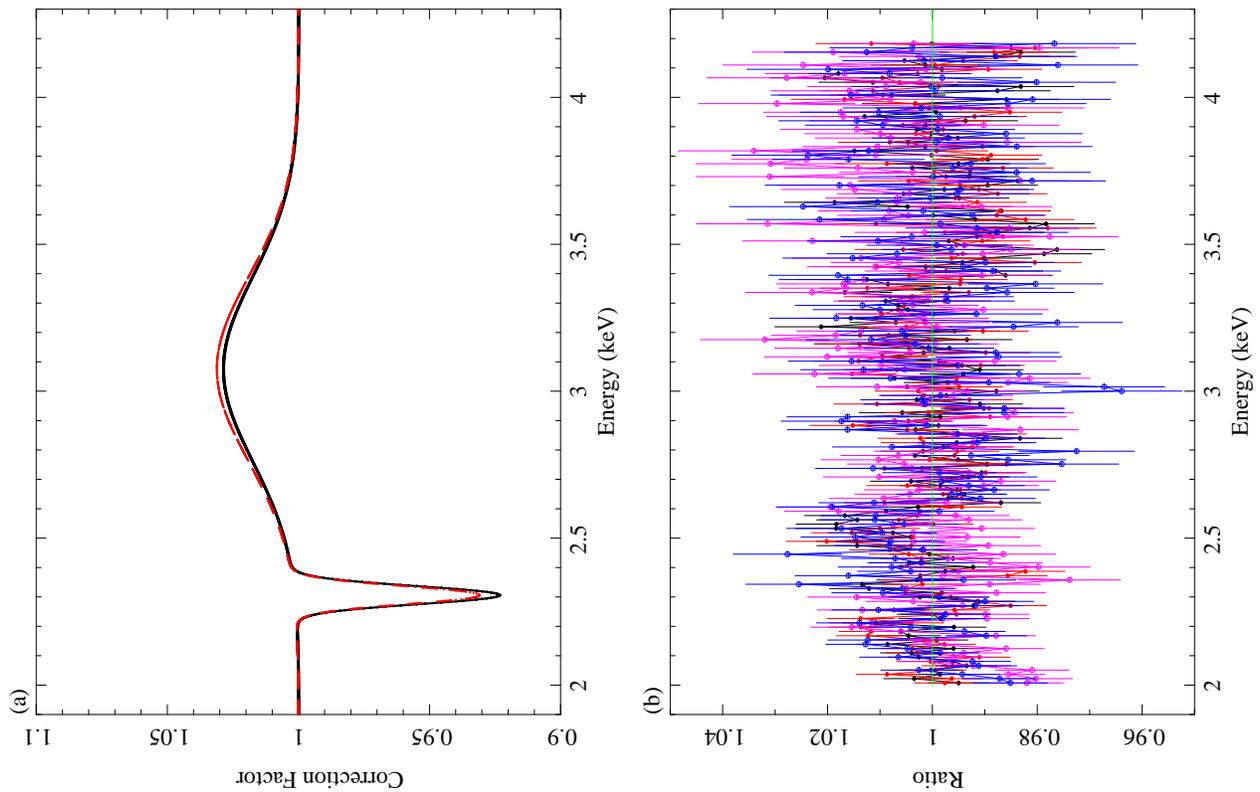

\begin{center}
\includegraphics[scale=.44]{fig-crab-42b.ps}
\includegraphics[scale=.44]{fig-crab-42b-ratio.ps}
\caption{
(a) Correction function from the 2.3~keV and 3.1~keV Gaussian modifications.
Black and red lines present FI and BI models, respectively.
(d) Fitting residuals in data to model ratio for the 'offset$+$2 Gaussians' model.
The data and model are the same as panel (a),
but with the correction given in the top panel.
Different marks represent different data sets.
}
\label{crab:fig:0827-41}
\end{center}
\end{figure}

We model these two residuals 
by assuming errors in the effective area and 
introducing
Gaussian multiplicative modifiers 
which enhance or attenuate the emission.
To represent the attenuated feature around 2.3~keV
and the enhanced one peaked around 3.2~keV, 
two 'gabs' components in XSPEC are used.
For each of these Gaussian components, 
three parameters are adjustable:
central energy, sigma, and strength.
There is no significant change in the residuals by observation period.
Furthermore, residuals in the FI and BI CCD spectra
are similar in shape.
Therefore, we assume a common correction for the two periods
and common parameters for the FI and BI CCD spectra 
as given in Table~\ref{crab:0827-41}.
The sigma for the first Gaussian is fixed at 0.03~keV
to avoid over-sharp modifications.
The best-fit correction parameters are given in Table~\ref{crab:0827-41}
and illustrated in Fig.~\ref{crab:fig:0827-41}.
Using these modifications in the response ('offset$+$2 Gaussians' in Table~\ref{crab:0827-41}),
we obtained 
significantly better fit 
($\Delta \chi^2=1072$ for seven additional parameters)
with residuals below a few percent over the 2--4.4~keV band.
We call these 2.3~keV and 3.1~keV Gaussian modifications.

\begin{table}
\caption{XIS Crab spectral fitting results.}
\label{crab:fit-para}
\begin{center}
\begin{tabular}{llr}
\hline
model name       & index\footnotemark[$\sharp$] & $\chi^2/d.o.f.$ \\
\hline
standard response               & 2.05      & 2020/595 \\ %
offset correction\footnotemark[$*$]            & 2.04     & 1778/591 \\ 
offset$+$2 Gaussians & 2.06 & 706/584 \\ 
\hline
\end{tabular}
\end{center}
\footnotemark[$*$] 
The offset are 6.2, 7.5, 11.9, and 6.7~eV, for 
FI/AO3-4-5, FI/AO6-8,
BI/AO3-4-5 and BI/AO6-8 spectra, respectively.
\footnotemark[$\sharp$]
Index errors are about $2\times 10^{-3}$.
\end{table}

\begin{table}
\caption{Correction parameters for the two Gaussian modifiers  from the XIS Crab spectral fitting.}
\label{crab:0827-41}.
  \begin{center}
\begin{tabular}{lllllll}
\hline
sensor & \multicolumn{3}{c}{1st Gaussian} & \multicolumn{3}{c}{2nd Gaussian} \\
       & energy & sigma & strength & energy & sigma & strength \\
       & (keV) & (keV) &               & (keV) & (keV) &               \\
\hline
FI     & 2.31 & 0.03\footnotemark[$*$] & $6.1\times 10^{-3}$ & 3.08 & 0.32 & $-2.30\times 10^{-2}$\\
BI     & $=$  &  $=$  & $5.5\times 10^{-3}$ & $=$  &  $=$ & $-2.43\times 10^{-2}$\\

\hline
\end{tabular}
  \end{center}
\footnotemark[$*$] This value is fixed.\\
'=' indicates the same value as in the FI model.
\end{table}

\subsection{Systematic Response Uncertainty}
\label{crab-limit}

\begin{figure}[hpt]
\begin{center}
\includegraphics[scale=.5]{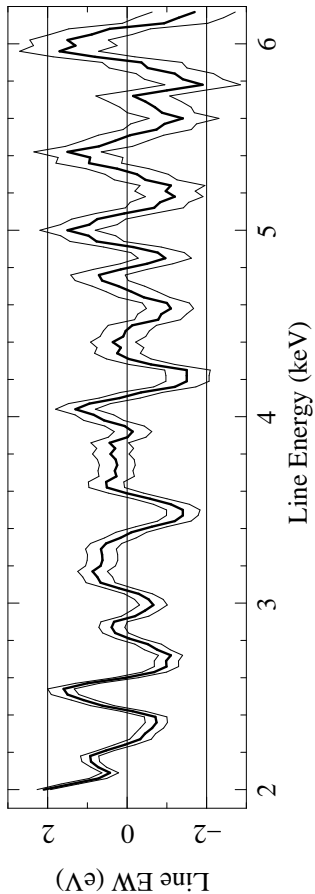}
\caption{
Results of the search for line emission or absorption
in the Crab spectra.
Three lines indicate the best-fit and 1$\sigma$ (68\%) confidence range 
in units of equivalent width as a function of given line center.
}
\label{crab:0828-1}
\end{center}
\end{figure}

We estimate the systematic sensitivity limited by the effective area calibration in detecting a line feature.
We add a Gaussian line emission or absorption component to fit the Crab spectra
and determine the best-fit and confidence range of the line flux
along with the change in $\chi^2$
for a grid of fixed line centers.
For this fit,
we localized the energy range within 400~eV (about four times the full width half maximum of the energy resolution)
centered on the given energy.
We applied the same model, gain tuning, and two Gaussian modifications ('offset$+$2 Gaussians') as obtained in the previous subsection.
Only normalizations of the power-law and the Gaussian line are allowed to vary.
Thus the line normalization obtained is converted into line equivalent width (EW) and given in Fig.~\ref{crab:0828-1}.

The EW obtained values fluctuate not only statistically 
but also systematically due to response errors.
At some energies, corresponding to residuals in the Crab fit (Fig.~\ref{crab:fig:0827-41}),
false signals in line emission or absorption are detected.
The best-fit values of $|EW|$ are less than 2~eV.
This EW profile estimates the systematic sensitivity
for detection of a weak line feature, limited by the energy response calibration.

\section{\suzaku Observations and Data Analysis}
\subsection{Observations and Basic Data Reduction}
The Perseus cluster center has been observed twice-yearly 
as a calibration target of the XIS
and we have used all available data.
These data were used in the instrument calibration reported by \citet{koyama07}, \citet{ozawa09}, and \citet{uchiyama09}
and in cluster spectroscopic studies (e.g. \cite{tamura2014}).

We used the same data and analysis method as those in \citet{tamura2014} and \citet{t09}
and added new observations.
Data obtained from 2007 to 2014 taken in the normal window and SCI 'on' mode
are used as shown in Table~\ref{obs:obs-center}.
The total exposure time is more than 500~ks for each CCD as given in Table~\ref{obs:xis-epic}.
We use XIS~0 (FI), XIS~1 (BI), and XIS~3 (FI) (XIS~2 is not used).

Detailed descriptions of the \suzaku observatory, XIS instrument, 
and X-ray telescope  are found in \citet{mitsuda07}, \citet{koyama07}, and 
\citet{serlemitsos07}, respectively.

We used the latest calibration file as of February 2014.
For spectra fitting, we used XSPEC and 
$\chi^2$ statistics implemented in this package.

We compare our \suzaku XIS observations
with the \xmm EPIC ones used in B14 (Table~\ref{obs:xis-epic}).
Because of the longer XIS exposures,
the entire \suzaku observations should collect
at least four times as many photons
from the same solid angle as do the \xmm observations.
Even taking into account that the \xmm has twice as large a field of view,
and assuming uniform photon distribution over the field of view, 
the XIS collects a larger number of photons than the EPIC.

\begin{table}
\caption{
\suzaku central pointings of the Perseus cluster. 
}
\label{obs:obs-center}
\begin{center}
\begin{tabular}{llrrl}
\hline
Date & Sequence    & 
Exp (a) & 
Roll (b) & Note \\
       & & (ks) & (degree) & \\
\hline
2007 Feb & 101012020 & 40.0 & 258.7 &  \\
2007 Aug & 102011010 &  35.1& 83.4 & \\
2008 Feb & 102012010 &  34.9 & 255.2 & \\
2008 Aug & 103004010 & 34.1 & 86.8  &\\
2009 Feb & 103004020 & 46.3 & 256.1 & \\
2009 Aug & 104018010 & 34.2 & 67.0 &  \\
2010 Feb & 104019010  & 33.6 & 277.3 &  \\

2010 Aug & 105009010 & 29.6 & 66.6  & \\
2011 Feb & 105009020 & 32.9 & 259.7 &  \\

2011 Aug & 106005010 & 34.1 & 83.8 &   (c) \\
2012 Feb & 106005020 & 41.1 & 262.0  & \\

2012 Aug & 107005010 & 33.2& 72.6  & \\
2013 Feb & 107005020 & 35.6 & 256.3 & \\
2013 Aug & 108005010 & 38.1& 76.2 & \\
2014 Feb & 108005020 & 34.1& 256.0 & \\
\hline
\end{tabular}
\end{center}
(a) Exposure time.\\
(b) Roll angle of the pointing defined as north to DETY axis. \\
(c) No XIS-3 data is available.\\
\end{table}

\begin{table}
\begin{center}
\caption{\suzaku XIS and \xmm EPIC observations.
}
\label{obs:xis-epic}
\begin{tabular}{lllllll}
\hline
Detector & Area\footnotemark[$*$] & FOV\footnotemark[$\dagger$] & exp\footnotemark[$\sharp$] & Area$\times$ exp & Area$\times$ exp $\times$ FOV \\
         & (cm$^2$) & (arcmin$^2$) & (ks) & 
($10^6$ cm$^2 \cdot $ s) & 
($10^9$ cm$^2 \cdot $ s $\cdot$ arcmin$^2$) \\
\hline
XIS/FI & 260 & 320 & 1040 & 270 & 86.5 \\
XIS/BI & 260 & 320 & 530  & 138 & 44.1 \\
total & -    & -   & -    & 408 & 130.6 \\
\hline
MOS & 300 & 710 & 317 & 95.1 & 67.5 \\
pn  & 700 & 710 & 38  & 26.6 & 18.9 \\
\hline
\end{tabular}
\end{center}
\footnotemark[$\dagger$] Detector's field of view.\\
\footnotemark[$*$] Effective area at energy of 3.5~keV.\\
\footnotemark[$\sharp$] Exposure time. EPIC values are those of \cite{bulbul2014}.
The FI and MOS exposures are the sums of those of each sensor (i.e., XIS-0$+$XIS-3 or MOS-1$+$MOS-2).
\end{table}

\subsection{Spectral Extraction and Responses}

\label{sec:ana}
We extracted spectra from three annular regions with boundary radii of $0'.5$, $2'$, $4'$, and $10'$ 
and one circle with radius of $10'$, all centered on the X-ray maximum.
We call these C05-2, C2-4,C4-10, and C10, respectively.
Similar regions are used in \citet{t09}.
Note that C10 overlaps with other regions.

Spectra from all observations are combined for each FI (XIS~0 and XIS~3) and BI CCD.

\begin{figure}[hpt]
\begin{center}
\includegraphics[scale=.7]{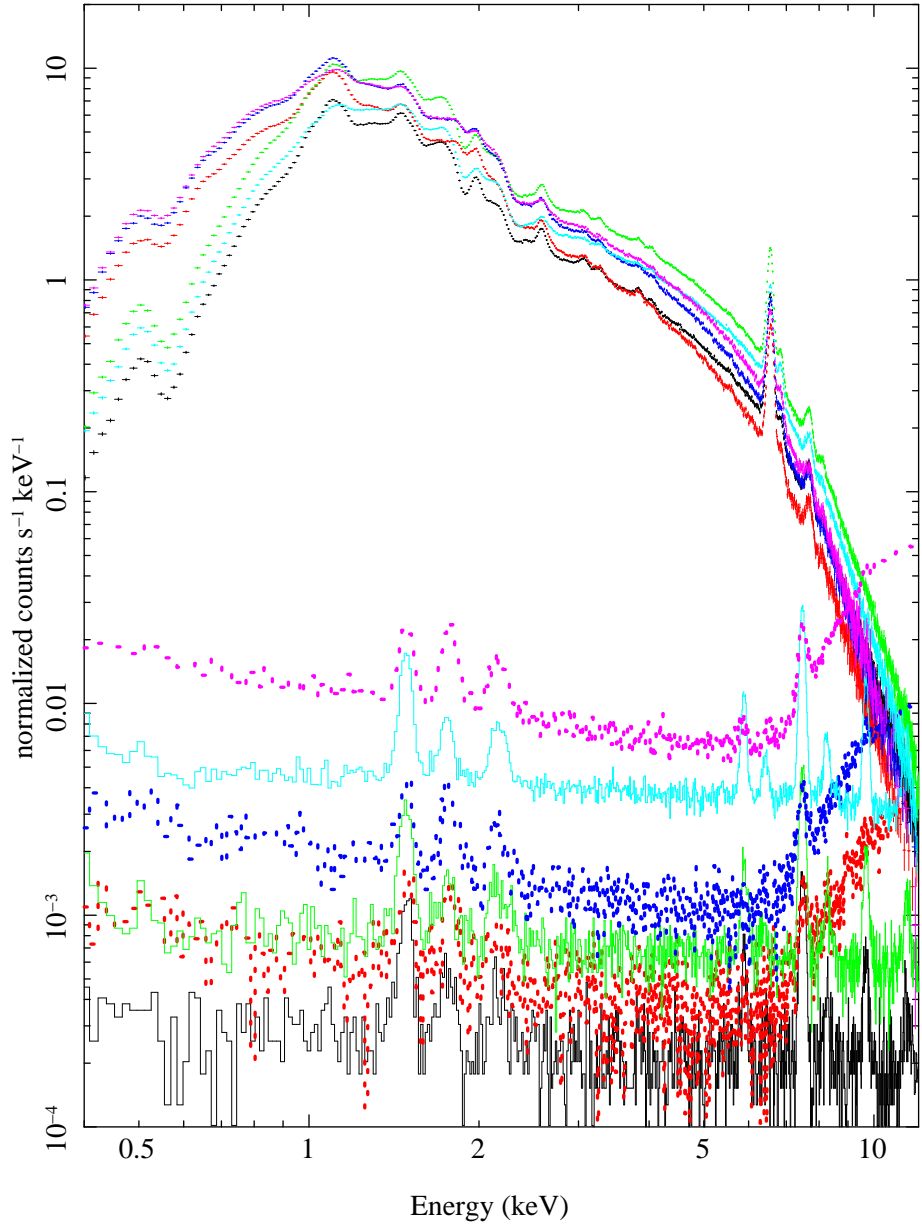}
\caption{
The XIS spectra extracted from the Perseus central pointings
are shown.
Six spectra from FI and BI CCD and from three annular regions with boundary radii of 
$0'.5-2'$ (black for FI, red for BI), 
$2'-4'$ (green and blue),
and $4'-10'$ (light blue and magenta)
are shown along with estimated instrumental background (in the same colors).
}
\label{ana:rall}
\end{center}
\end{figure}

The instrumental (non-X-ray) background was estimated using the night earth observation database and the {\tt xisnxbgen} software \citep{tawa2008}.
We extracted this background from the source before spectral fitting.
The cosmic X-ray background was estimated to be well below 1\% of the source over almost the entire energy band.
We therefore ignore this cosmic background in the following analysis. 
Fig.~\ref{ana:rall} shows extracted source and instrumental background spectra.

We use the same ARFs as used in the Crab fitting in \S~\ref{crab-obs}.
We produced RMF files for each observation period.
Among them, 
we found the AO-3 period file set best reproduces the data.
Therefore we used this RMF set for the spectral fitting.
The energy resolution at around 3.5~keV of the XIS FI spectra
is about 100~eV (full width half maximum).
Therefore no information is lost if a bin size below 100/3 ~eV is used.
We rebin spectra into four bins (energy bin of 3.65~eV).

\subsection{Local Band Fitting}
\label{fit-local}

Here we inspect spectra for the 3.5~keV line emission suggested 
by B14 ($3.57\pm 0.02$~keV in the cluster's rest frame).
Table~\ref{ana:line-ene} lists measured line positions alongside other relevant transitions.
We fitted FI and BI spectra in the 2.8--4.2~keV energy band including He- and H-like Ar and Ca lines
with the following two models.
One model, 'VAPEC' in XSPEC
describes the thermal emission from a collisional ionization equilibrium plasma \citep{smith01}.
The solar metal abundances are taken from \citet{ag89}.
Free parameters are temperature ($kT$), 
metal abundances of Si, S, Ar, Ca, Fe (equivalent to Ni), redshift, and normalization.
Other metal abundances including Cl and K are fixed to 1.0 by setting the XSPEC parameter {\tt APEC\_TRACE\_ABUND} to be 1.0.
The other model is a 'line-free APEC' continuum plus some Gaussian lines ('Line0APEC+Lines').
The line-free APEC (Astrophysical Plasma Emission Code) model represents continuum emission
accounting for thermal bremsstrahlung, 
radiative recombination,
and two-photon emission, 
and was used in B14.
We use a metal abundance of 0.5 taken from \citet{t09}.
Free parameters are $kT$, normalization of the line-free APEC,
normalization of each Gaussian line, 
and a common redshift.
We include some transitions from He-like and H-like ions within the fitting band given in Table~\ref{ana:line-ene}.

For these fittings we do not apply 'gain tuning' or the Gaussian modifications as used in \S~\ref{crab-obs}.

The fitting results are shown in 
Table~\ref{fit-local:fit}
and Figs.~\ref{fit-local:fit1} and ~\ref{fit-local:fit2}.
Either model describes these spectra accurately with residuals of only a few percent.
In all cases, 
four lines from Ar and Ca, 
but no line-like structure around 3.5~keV between these lines, 
are seen.
To demonstrate the lack of line feature at the energy, 
we compare the data with a model with the line feature with the flux measured by B14
in Fig.~\ref{fit-local:fit2-1211}.

\begin{table}
\begin{center}
\caption{Local-band spectral fitting results for the 2.8-4.2~keV band.
}
\label{fit-local:fit}
\begin{tabular}{llllll}
\hline
Region & $\chi^2/$d.o.f. & $kT$ & Ar & Ca & redshift \\
       &                 & (keV) & (solar) & (solar) & \\
\hline
\multicolumn{6}{c}{VAPEC} \\
C05-2 & 215/180 & 3.6 & 0.67 & 0.73 & 0.015 \\
C2-4  & 229/180 & 3.9 & 0.71 & 0.65 & 0.014 \\
C4-10 & 237/180 & 5.5 & 0.53 & 0.54 & 0.016 \\
C0-10 & 361/180 & 3.9 & 0.71 & 0.66 & 0.015 \\ 
\hline
\multicolumn{6}{c}{Line0APEC$+$Lines } \\
C05-2 & 225/190 & 4.3 & - & - & - \\
C2-4  & 197/190 & 5.0 & - & - & - \\
C4-10 & 234/190 & 5.9 & - & - & - \\
C0-10 & 292/190 & 5.0 & - & - & - \\
\hline
\end{tabular}
\end{center}
Statistical errors of $kT$, Ar, Ca, and redshift
are about
0.05-0.1~keV
0.05-0.1 solar,
0.05-0.1 solar,
and $0.0002-0.0004$.
\end{table}

\begin{figure}[hpt]
\begin{center}
\centerline{\hbox{
\graphicspath{{/Users/ttamura/perseus/35kevline/paper2014/}}
\includegraphics[scale=.38]{fig-vapec-r05-2.ps}
\includegraphics[scale=.38]{fig-nolineGau10-2-r05-2.ps}
}}
\centerline{\hbox{
\includegraphics[scale=.38]{fig-vapec-r2-4.ps}
\includegraphics[scale=.38]{fig-nolineGau10-2-r2-4.ps}
}}
\centerline{\hbox{
\includegraphics[scale=.38]{fig-vapec-r4-10.ps}
\includegraphics[scale=.38]{fig-nolineGau10-2-r4-10.ps}
}}
\caption{
The XIS FI (black) and BI (red) spectra 
fitted with 
VAPEC (left) and Line0APEC$+$Lines (right) models.
Annular regions are specified in each panel.
Vertical axes for each panel are counts s$^{-1}$ keV$^{-1}$
and date to model ratio.
}
\label{fit-local:fit1}
\end{center}
\end{figure}

\begin{figure}[hpt]
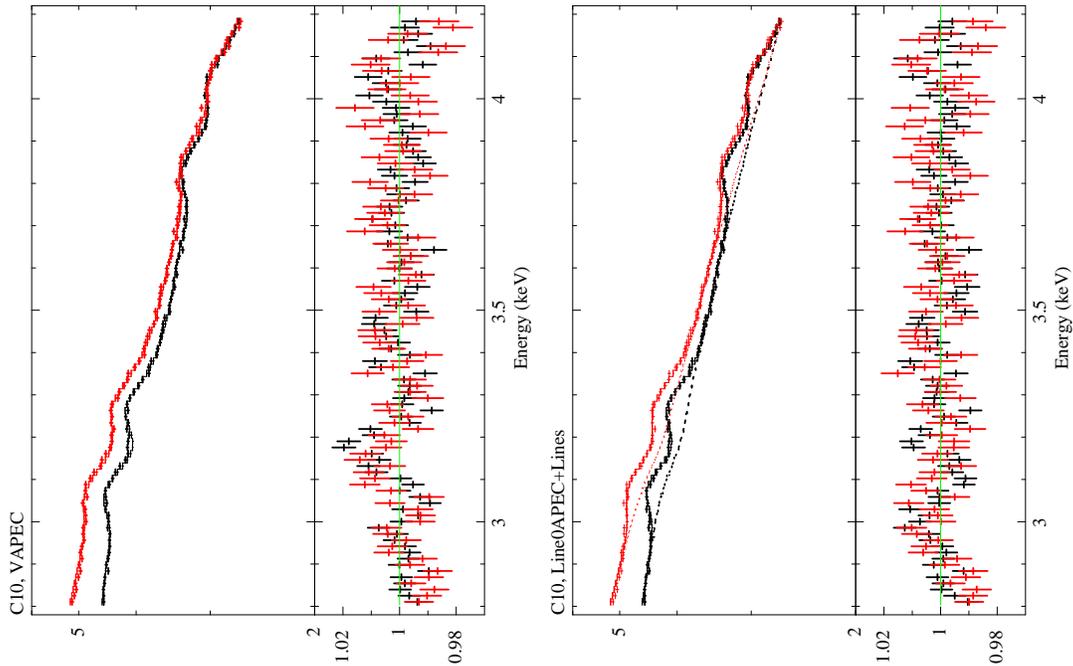

\begin{center}
\centerline{\hbox{
\includegraphics[scale=.38]{fig-vapec-r10.ps}
\includegraphics[scale=.38]{fig-nolineGau10-2-r10.ps}
}}
\caption{
Same as the previous plot, but for C10 annular region spectra.
}
\label{fit-local:fit2}
\end{center}
\end{figure}

\begin{figure}[hpt]
\begin{center}
\includegraphics[scale=.38]{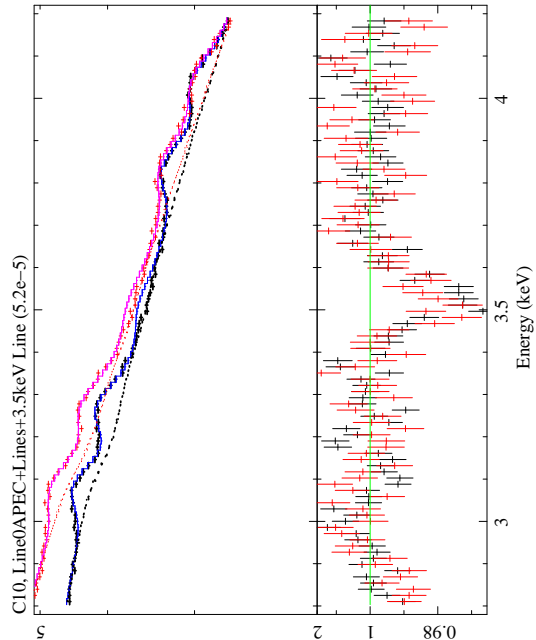}
\caption{
Same as the previous plot, 
but with model (blue and magenta) including a line emission at 3.51~keV with a flux measured in B14 
($5.2 \times 10^{-5}$ photons cm$^{-2}$ s$^{-1}$).
}
\label{fit-local:fit2-1211}
\end{center}
\end{figure}

\begin{table}
\begin{center}
\caption{Relevant line positions}
\label{ana:line-ene}
\begin{tabular}{rlll}
\hline
Energy & Ion & Type and notes\\
(keV)  &     &  \\
\hline
\multicolumn{4}{c}{Unidentified line measured in \cite{bulbul2014} (Table 5)} \\
\multicolumn{4}{c}{All in rest frame energy} \\
$3.57\pm 0.02$ & & from the MOS \footnotemark[$\dagger$] \\
$3.51\pm 0.03$ & & from the pn \\
$3.56\pm 0.02$ & & from the ACIS-S \\
\hline
\multicolumn{4}{c}{Used in 'Line0APEC+Lines' model (\S~\ref{fit-local},~\ref{sect:fit-limit}) \footnotemark[$\sharp$]} \\
2.0050 & Si XIV & H-like, Ly$\alpha$ \\
2.4600 & S XV & He-like, triplet \\
2.6220 & S XVI & H-like,  Ly$\alpha$ \\
3.1400 & Ar XVII & He-like, triplet \\
3.3230 & Ar XVIII & H-like, Ly$\alpha$ \\
3.9020 & Ca XIX & He-like, triplet \\
4.1080 & Ca XX & H-like, Ly$\alpha$ \\
5.6820 & Cr XXIII & He-like, triplet \\
6.7000 & Fe XXV & He-like, triplet \\
\hline
\multicolumn{4}{c}{Used  in \citet{bulbul2014} within 3--6~keV } \\
3.12 & Ar XVII &  \footnotemark[$\ddagger$]\\
3.31 & Ar XVIII & \footnotemark[$\ddagger$]\\
3.47 & K XVIII & \\
3.51 & K XVIII & \\
3.62 & Ar XVII & DR [$*$] \\
3.68 & Ar XVII & \\
3.71 & K XIX  & \\
3.86 & Ca XIX &  \\
3.90 & Ca XIX & \footnotemark[$\ddagger$]\\
3.93 & Ar XVIII & \\
4.10 & Ca XX & \footnotemark[$\ddagger$]\\
4.58 & Ca XIX & \\
5.69 & Cr XXIII&\\
\hline
\end{tabular}
\end{center}
\footnotemark[$\dagger$] at the Perseus $z$, observed at $3.57/1.016=3.51$~keV \\
\footnotemark[$*$]dielectronic recombination \\
\footnotemark[$\sharp$]
All positions are taken from the AtomDB database.
\footnotemark[$\ddagger$] Correponding transitions are given in the second rows, 
``Used in 'Line0APEC+Lines' model''.
\end{table}

\subsection{Line Flux Limits}
\label{sect:fit-limit}

\subsubsection{Template model}
We search the spectra for line features not only at 3.5~keV but also in a range of line positions.
We use C10 spectra extracted from the largest region to obtain the best statistics
and use the spectra in the 2.0--6.3~keV range to obtain better continuum emission models.

We determine a spectral model
having physically reasonable parameters
and describing the data accurately.
To do this, we apply several corrections to the response and model, as follows.
We found that 
neither the energy scale offset correction
nor the 2.3~keV modification as used in \S~\ref{crab-obs}
improves the fit.
Therefore we apply only the 3.1~keV Gaussian modification.
We use the same 'Line0APEC$+$Lines' model as above
by adding other transitions within the wider energy band (Table~\ref{ana:line-ene}).
We also add the He-like Fe-K transition (at 6.7~keV)
to model the Si-escape feature of this transition 
as seen in the spectra around 4.8~keV.
To compensate for uncertainties
of line central energies in the APEC model
and 
of the instrumental energy-scale,
we allow redshift for each line
to vary within $\pm$0.5\% of the model value.
 ($\Delta z < 0.005$).

\begin{figure}[pth]
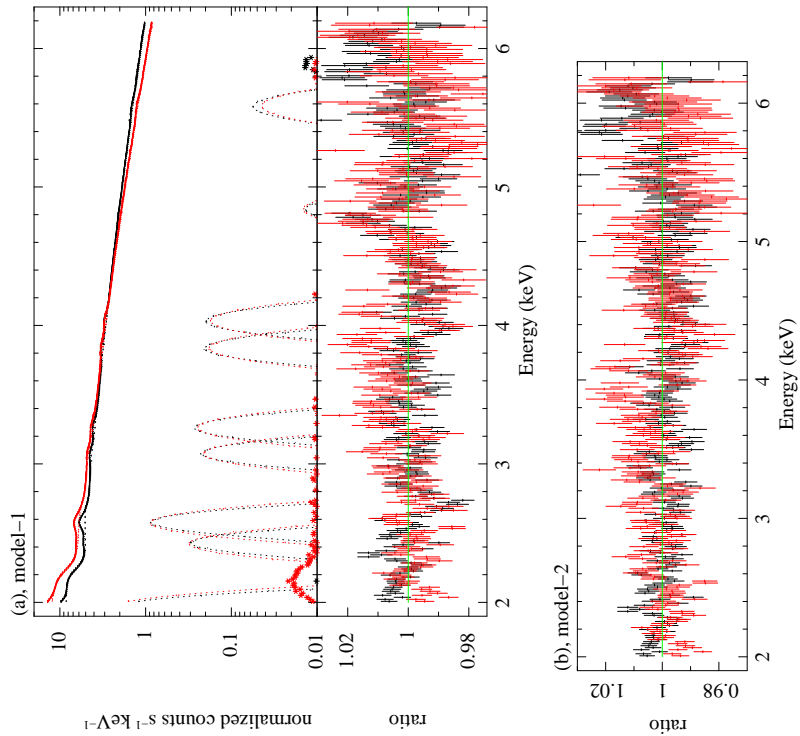

\begin{center}
\includegraphics[scale=.38]{fig-0917-1fit.ps}
\includegraphics[scale=.38]{fig-0924-2-nofit-ratio.ps}
\caption{
(a) The Perseus central spectra (C10) fitted with 'model-1'.
Background spectra are also shown in the upper panel.
Black and red colors present the FI and BI data respectively.
(b) 
Plots as panel (a),  but with an additional three line components in the model (model-2).
}
\label{fit-limit:model12}
\end{center}
\end{figure}

The fitting results are shown in Fig.~\ref{fit-limit:model12}(a).
This model ('model-1') provides a good fit with a $\chi^2$ value of 1127 for 561 degrees of freedom. 

\begin{figure}[hpt]
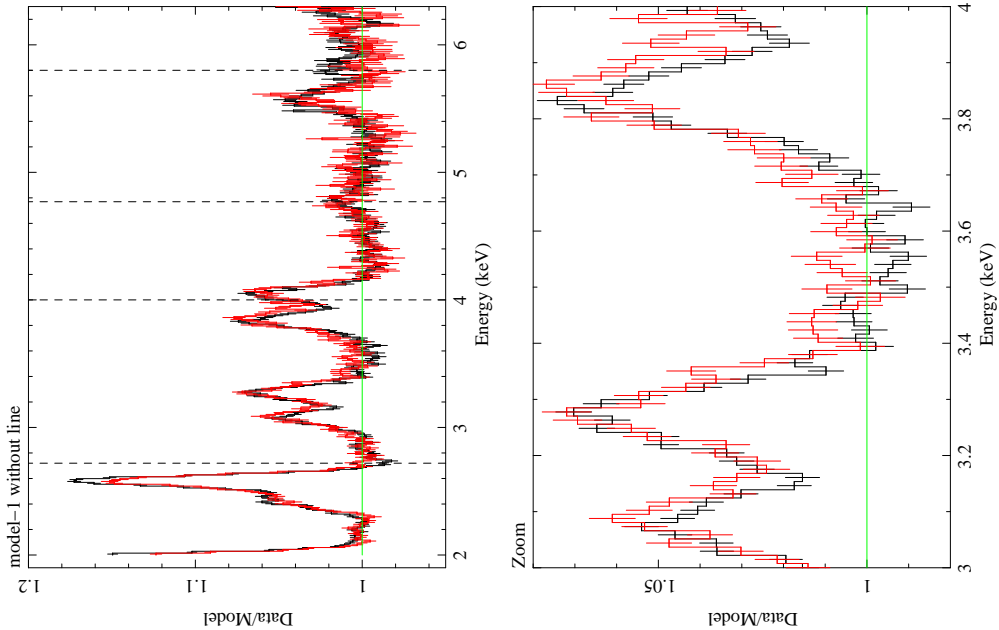

\begin{center}
\centerline{\hbox{
\includegraphics[scale=.35]{fig-0917-3ratio.ps}
\includegraphics[scale=.35]{fig-0917-3ratio2.ps}
}}
\vspace{1cm}
\caption{
The data ratio to the model ('model-1') without line emission component.
Right panel shows a zoom into the energy range of interest.
 }
\label{fit-limit:ratio}
\end{center}
\end{figure}

We show line emission features
in terms of the data ratio to the same model but without line emission (Fig.~\ref{fit-limit:ratio}).
This clearly demonstrates
detection of known atomic lines
and a lack of line features 
between the Ar and Ca lines in energy.

There are some residuals of a few percent of the observed counts.
Among them,
we identified four systematic features as listed in Table~\ref{fit-limit:res} and marked in Fig.~\ref{fit-limit:ratio}.
The first feature is also seen in the Crab spectral fitting (see e.g. Fig.~\ref{crab:0828-1})
and related to an instrumental Au edge.
The second is between He- and H-like Ca transitions.
The third is close to the Fe-K Si escape line.
To model these three residuals 
we added three Gaussian emission (for the third residual) or absorption (for the first and second) components.
This new model ('model-2') provides better fit, as expected [Fig.~\ref{fit-limit:model12}(b)].
The fitting $\chi ^2$ is 959 for 558 degrees of freedom.
We do not attempt to model the fourth background residual.

\begin{table}
\begin{center}
\caption{
Residual features identified in the C10 spectral fitting in the 2.0--6.3~keV band. 
}
\label{fit-limit:res}
\begin{tabular}{lllll}
\hline
\# & (1) & (2) apparent position & (3) EW & related feature \\
   &     & (keV)    &   (eV)     & \\
\hline
1 & $-$ & 2.72 & 1.4 & Au M3 edge at 2.743~keV \\
2 & $-$ & 4.00 & 1.6 & He- and H-like Ca \\
3 & $+$ & 4.77 & 2.5 & Fe-K Si escape at 4.85~keV \\
4 & $+$ & (5.8) & -  & Mn-K$\alpha$ at 5.90~keV \\
\hline
\end{tabular}
\end{center}
(1) $+$ and $-$ indicate excess and attenuation features, respectively.\\
(2) Line positions in the fitting model ('model-2').\\
(3) Estimated line strength in EW. 
\end{table}

We also model spectra in the $E<2.0$~keV and $E>6.2$~keV bands, 
to be discussed in the next subsection.
In these cases, we use the VAPEC model.
In the lower energy band  there are known response problems with the detector, for example around the Si edge.
In the higher energy band residuals from the instrumental background can be found.
These instrumental features result in systematic residuals up to 5--10\% in the data-to-model ratio.
The fitting results are given in the appendix.


\subsubsection{Line flux limit}
\label{sect:line-limit}

\begin{figure}[hpt]
\begin{center}
\includegraphics[scale=.50]{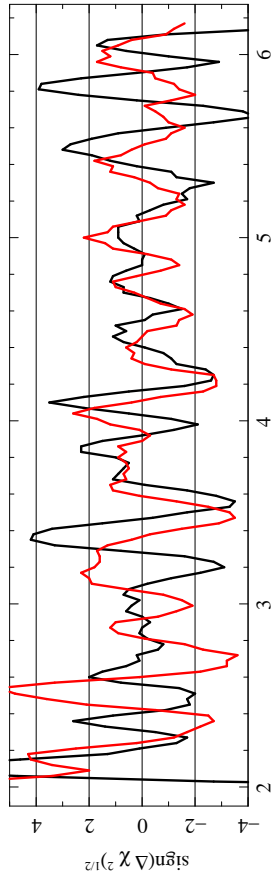}
\caption{
The weak line search result with the Perseus C10 spectra (black) 
and the Crab spectra (dotted and red).
This plot shows ($\Delta \chi^2)^{1/2}$ (in unit of $\sigma$) 
between the local models with and without line feature at the energy given on the horizontal axis
with positive (emission) or negative (absorption) signs in the best-fit line intensity.
}
\label{fit-limit2:sdelta}
\end{center}
\end{figure}

\begin{figure}[hpt]
\begin{center}
\includegraphics[scale=.60]{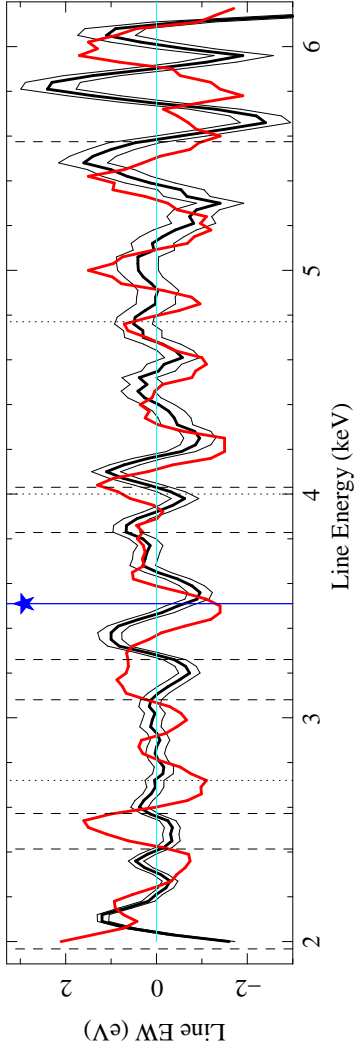} 
\caption{
The same result as Fig.~\ref{fit-limit2:sdelta}, 
shown in terms of the best-fit line intensity in EW
and its 68\% confidence ranges.
Positions of atomic lines from Si, S, Ar, Ca, Mn
and the identified features (2.72, 4.00, and 4.77~keV; Table~\ref{fit-limit:res})
are shown by dashed and dotted lines, respectively.
}
\label{fit-limit2:e-ew}
\end{center}
\end{figure}

Based on the model correcting the three residuals obtained above ('model-2')
we searched weak line features by the same method as used in \S~\ref{crab-limit}.
Results of the change in $\chi^2$ and line EW for the energy range 
are given in Figs.~\ref{fit-limit2:sdelta}-\ref{fit-limit2:e-ew}.
We compare the EW limits with those for the Crab spectra (Fig.~\ref{crab:0828-1}).
Best fit line flux values 
vary over energies in positive (emission)
and negative (absorption) intensities 
due to systematic residuals and statistical fluctuations.
Large features are seen close to the lowest bound at 2~keV. 
There are also large fluctuations above 5.5~keV including 
instrumental background line features identified above (right column in Table~\ref{fit-limit:res}).
At energy ranges around the three identified and modeled features (Table~\ref{fit-limit:res}), 
line intensity errors could be larger than those given in Fig.\ref{fit-limit2:e-ew}.
Other fluctuations are mostly less than $\pm$1~eV in EW
and not significantly larger than systematic features seen in the Crab fit.
Therefore we conclude that at least in the energy range of 2--5.5~keV, 
except for ranges close to the three residuals, 
no line emission stronger than 1~eV in EW was detected.

We compare our limit with the detected line flux of the unidentified 3.5~keV emission in B14.
Taking continuum flux and effective area from their Fig.5, 
we translate their flux of 
$(5.2 ^{+3.7}_{-2.1}) \times 10^{-5}$ photons cm$^{-2}$ s$^{-1}$ 
(90\% confidence limits)
for the Perseus cluster (with the core)
into a line EW of 2.9~eV , shown in Fig.~\ref{fit-limit2:e-ew}.
Their detected position of $3.57\pm 0.03$~keV (90\% confidence) in the rest-frame
is redshifted to $3.51\pm 0.03$~keV.
At this position,
our limit of the line EW from the Perseus spectra is 
-(1.0--0.5)~eV
, well below the \xmm best-fit value.
At this energy the Crab spectral fitting gives line fluxes in minus EW down to -1.5~eV.
Accordingly we estimate the systematic line flux uncertainty around this energy
to be 1.5~eV 
(68\% confidence limit).
By scaling the line flux in B14 given above by the EW ratio
this EW corresponds to a line flux
of 
$5.2 \times \frac{1.5}{2.9} \times 10^{-5} = 2.7 \times 10^{-5} $ photons cm$^{-2}$ s$^{-1}$.
This estimate is a half of the {\it XMM-Newton}-detected flux but still close enough to be within 90\% confidence of the {\it XMM-Newton}-detected flux.

\subsection{Line Flux Measurements of the Unidentified Line at 3.5~keV}

\begin{figure}[pth]
\begin{center}
\includegraphics[scale=.40]{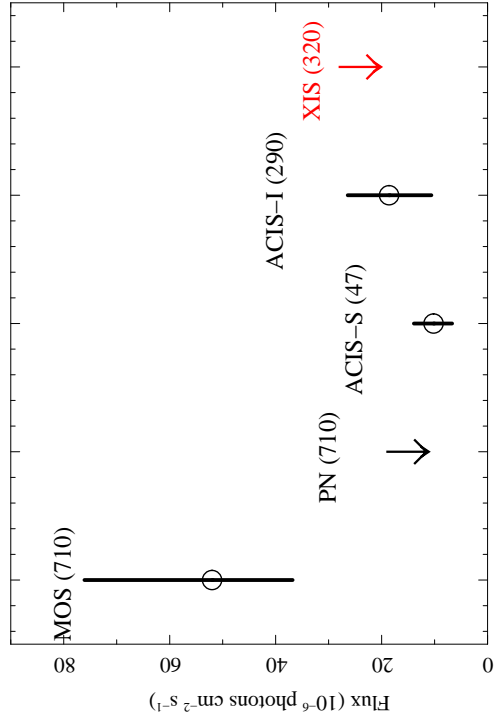}
\caption{
Line fluxes in the Perseus center from 
\xmm, \chandra and \suzaku detectors at 3.51~keV.
The \xmm and \chandra values are taken from \citet{bulbul2014} 
and include 68\% confidence statistical errors.
Numbers in parentheses after the detector name are sizes of spectral extraction in arcmin$^2$.
Note that pn and XIS spectra provided only an upper limit.
 }
\label{dis-com:flux_com}
\end{center}
\end{figure}

B14 detected an unidentified emission line
from the Perseus cluster center
at a rest-frame energy of $3.57\pm 0.02$~keV
with a flux of $5.2^{+2.4}_{-1.5} \times 10^{-5}$ photons cm$^{-2}$ s$^{-1}$ (68\% statistical confidence limits) with the \xmm MOS CCD.
They also measured line fluxes from the same object using other \xmm and \chandra detectors
as presented in Fig.~\ref{dis-com:flux_com}.
This MOS flux is the largest among measured fluxes from other clusters in B14.
To the contrary,
we found no sign of this line feature in the \suzaku XIS spectra 
from the same Perseus cluster center
around the same line energy
with a systematic error of 1.5~eV in equivalent width (\S~\ref{sect:fit-limit}).
We compared these line flux measurements in Fig.~\ref{dis-com:flux_com}.

For the 3.5~keV line flux, the detections in B14 
are inconsistent with our \suzaku limit, 
at least in the statistical sense.
In addition, B14 claimed detections
not only from the Perseus MOS spectra
but also from other detectors and other objects.
Therefore
the inconsistency should be explained not solely by statistical fluctuations
but also by systematic causes.

B14 discussed origins of the detected 3.5~keV line
by known atomic lines 
but with unexpectedly-strong emissivities 
or unexpected physical conditions to be assumed
(see section 5.1 in B14).
Any origins intrinsic to the Perseus cluster
can not explain the inconsistent results between this study and that in B14
since the two studies extracted spectra from almost the same region of the sky. 

\section{Discussion and Summary}

\subsection{Limitations of the \suzaku Line Search in the Perseus Spectra}
\label{sect:dis}
We discuss factors limiting the current line search with the \suzaku XIS in the bright Perseus cluster spectra.
First, 
we identify instrumental effects, 
which depend largely on line energies.
At lower energies ($E$) below 1~keV, 
the CCD efficiency drops alongside the strong O K-edge absorption.
In addition, time- and CCD position-dependent 
contamination in the CCD optical blocking filter complicates the energy responses.
Around the Si K-edge (1.6--1.95~keV)
there is a known systematic error in the CCD responses,
making line search difficult around this energy range. 
There is also a known error around an Au M-edge (2.2--2.4~keV; due to mirror)
of a level up to $\pm$ 5\% of the effective area.
In \S~\ref{sect:crab},
using the Crab spectra and standard responses,
we evaluated and corrected this 2.3~keV feature as much as possible, 
using a narrow Gaussian modification.
By introducing this and another modification,
the Crab spectra can be described in the 2--4.4~keV band 
within 2--3\% accuracy.
This accuracy, 
along with the Crab spectra above this range, 
indicates that
line detection sensitivity (limited by the energy response calibration)
is less than 2~eV in line equivalent width (EW) over the 2--6~keV band.
This sensitivity in most energy ranges is limited by statistical factors in the Crab observation data.
At $E>5.5$~keV, 
statistical and systematic uncertainties of 
instrumental background features
contribute to the line detection sensitivity.

Second, 
we identify factors
intrinsic to the Perseus spectra.
At $E<1$~keV, 
the cluster spectra contain the following complex structures: 
(1) cluster line emission from K-shell transitions of O and Ne and L-transitions of Fe, 
which is commonly strong in cool cluster cores such as the Perseus center; 
(2) soft X-ray background components from multiple origins (see e.g. \cite{Tamura2008}); 
(3) absorption due to the Galactic gas.
Further,  at 1.0~keV $<E<2.0$~keV, 
the cluster spectra
are crowded with lines from Ne-K, Mg-K, and S-K and the unresolved Fe-L complex.
These lines are also strong at metal-rich, cool cluster cores.
The CCD resolution is too low to detect a weak line within this cluster line emission.
To avoid these astronomical and instrumental features given the above considerations at low energies, 
we mainly focused on spectra above 2~keV.

Third, there is a statistical limit.
At 2.0~keV$<E<6.0$~keV,
the statistical uncertainty of the current Perseus observations
reaches a similar level (2~eV in EW) to the systematic uncertainty described above.

\subsection{A Possible Cause of Large Line Flux in \citet{bulbul2014}}
\label{dis-com2}

One clear difference between B14 and our \suzaku measurements
is in spectral extraction regions.
B14 collected spectra from the entire MOS field of view (FOV; $\sim 15'$~arcmin in radius $\sim 710$ arcmin$^{2}$),
while we used spectra from the XIS FOV (C10; $18'\times 18' \sim 320$ arcmin$^2$).
This cannot solely explain the difference in the line EW.
In fact, B14 found that more than $3/5$ of the total line flux
originates from the central $1'$ radius.
Therefore we can assume that the 3.5~keV line emission 
is not extended more than the cluster continuum emission.
This means that the EW within the XIS FOV should not be significantly smaller than
that within the larger MOS FOV.
The MOS vignetting also reduces effective area in detector outer regions.

Another difference is the modeling of weak atomic lines and continuum emission around the line position.
B14 already warned about effects of other line fluxes (in their section 5)
and continuum modeling (in their section 6).
B14 have examined these systematics using the Perseus \chandra spectra in their section 4.2.
However, because of the low statistical significance of their \chandra data,
it is not obvious that the same argument holds for other detections.
Therefore we suggest that their spectral modeling intrinsically tends to create an artificial excess particularly at 3.5~keV
in combination with other systematic errors, 
as we will argue below.

We included lines from H-like Ly$\alpha$ and He-like triplets of S, Ar, and Ca in our model,
because these are clearly detected in the XIS spectra.
In addition to these lines,
B14 included other known lines as given in Table~\ref{ana:line-ene}.
They allowed all line fluxes to vary
more than those expected from their estimation based on the AtomDB model, 
up to a factor of 3 in some cases.
They also allowed each line center energy and intrinsic width 
to vary up to 5~eV and 1\% of the energy resolution
to account for uncertainties in instrumental responses and atomic database.
In fact, 
we notice that the energy range around 3.5~keV
is unique without line emission in their model
(except for some weak lines which are not separated from the 3.5~keV line) 
in the 3--4~keV energy band,  as seen in their Fig.12.
Because CCD resolutions are too low to resolve these lines,
the best-fit line normalization 
could be larger than the real value
if systematic errors are not fully taken into account for the fitting.
For example,
systematic error may produce line fluxes larger than the real value,
and those lines in turn may reduce the continuum normalization below the true one.
This then creates an artificial excess around positions without line emission in the model.
Note that even in the Perseus cluster which provides the best data statistically,  
the excess detected in B14 is less than a few percent above the continuum.
This could be caused by a reduction of a few percent in the continuum normalization.
This effect tends to yield larger line flux in stronger continuum spectra such as the Perseus center.

\begin{figure}[hpt]
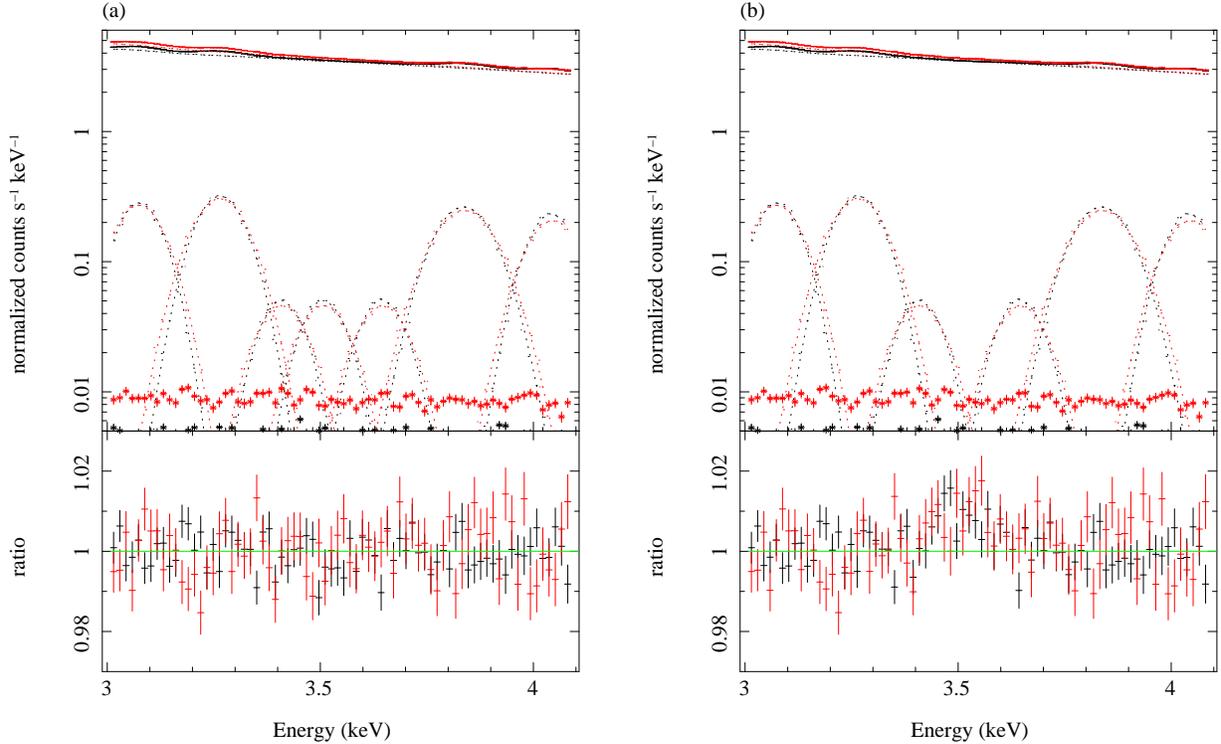

\begin{center}
\centerline{\hbox{
\includegraphics[scale=.40]{fig-1001-2a2.ps}
\includegraphics[scale=.40]{fig-1001-2b2.ps}
}}
\caption{
A simulation of the modeling effect of lines adjacent to the 3.5~keV line.
The same Perseus C10 XIS spectra are shown in both panels.
The FI and BI data are shown in black and red.
In panel (a) three lines are used to model the emission between the Ar and Ca lines.
In panel (b), the central line is removed from the model in (a).
A weak excess can be seen in the ratio.
}
\label{dis-com:1001-1}
\end{center}
\end{figure}

To demonstrate the effect given above quantitatively,
we use the same XIS spectra (C10) used in our analysis
and a model including additional lines as presented in Fig.~\ref{dis-com:1001-1}.
We use only the local energy range of 3--4.2~keV.
In addition to the four Ar and Ca lines,
we add new three line at 3.47, 3.57, and 3.71~keV, as used in B14.
These line normalizations are fixed to the same value, 
0.2 of that of He-like Ar line, 
which approximately corresponds to the MOS measured ratio in B14.
To allow maximum freedom, we allow central energy and width for each line to vary.
Because all energy bands are covered by unresolved line emission, 
it is difficult to set the normalization of the continuum.
Then, the continuum emission model (line-free APEC) are manually adjusted (i.e., parameters are not free).
By allowing the line parameter to vary, 
we can reproduce the XIS spectra with emission from the additional three lines 
as shown in left plot of the figure.
In the right panel,
we show the same data and model but without the 3.57~keV line 
illustrating a weak excess in the data-to-model ratio.

As given above, 
a weak line detection such as the 3.5~keV line, 
is sensitive to the modeling of continuum and neighboring unresolved lines particularly in low resolution spectra.
We suggest that this effect could possibly cause artificial detections of the 3.5~keV line in the spectral analysis of B14.

\subsection{Instrumental Calibrations}
\label{dis-cal}

As described in \S~\ref{sect:crab-mot}, 
instrumental calibration error could cause a false line detection.
In fact, 
there are features around 3.5~keV
in the XIS spectra (Fig.~\ref{ana-cal:arf})
and 
in the MOS and pn spectra (see Fig.7 of B14).
These features are probably related to an weak Au M edge
in mirror coatings in both \suzaku and \xmm.
We have evaluated this uncertainty
using the Crab XIS spectra
and found it to be less than 2~eV in EW around the line energy.
If the MOS has a similar level of error (~2~eV in EW) at the same energy position as 
but in a direction opposite to that in the XIS,  
the inconsistency between the two detectors can be completely removed.

B14 also commented briefly on this issue in their section 6.
They suggested that this error can be smoothed out and reduced for their stacked cluster sample analysis.
We further note that 
the \xmm MOS and pn systems have energy-dependent systematic errors on the effective areas
with a level of up to 5--10\% in the 2--5~keV band, 
as shown in \cite{Read2014} (see their Figs 1--2).

B14 also claimed a detection at the same energy in the Perseus cluster
with \chandra,
which uses a mirror without Au coating.
This detection, however, is less significant
and the measured flux is about five times smaller than the MOS result (see Fig.~\ref{dis-com:flux_com}).
More careful calibration and deeper exposures
are required to exclude the instrumental issues.

\subsection{Limits on Dark Matter Parameters}

Following previous studies (e.g. \cite{Abazajian2001} ),
we assume sterile neutrino decay X-ray and a dark matter profile, 
use the observed line flux limit over an energy range,
and constrain the dark matter decay rate.

\begin{figure}[hpt]
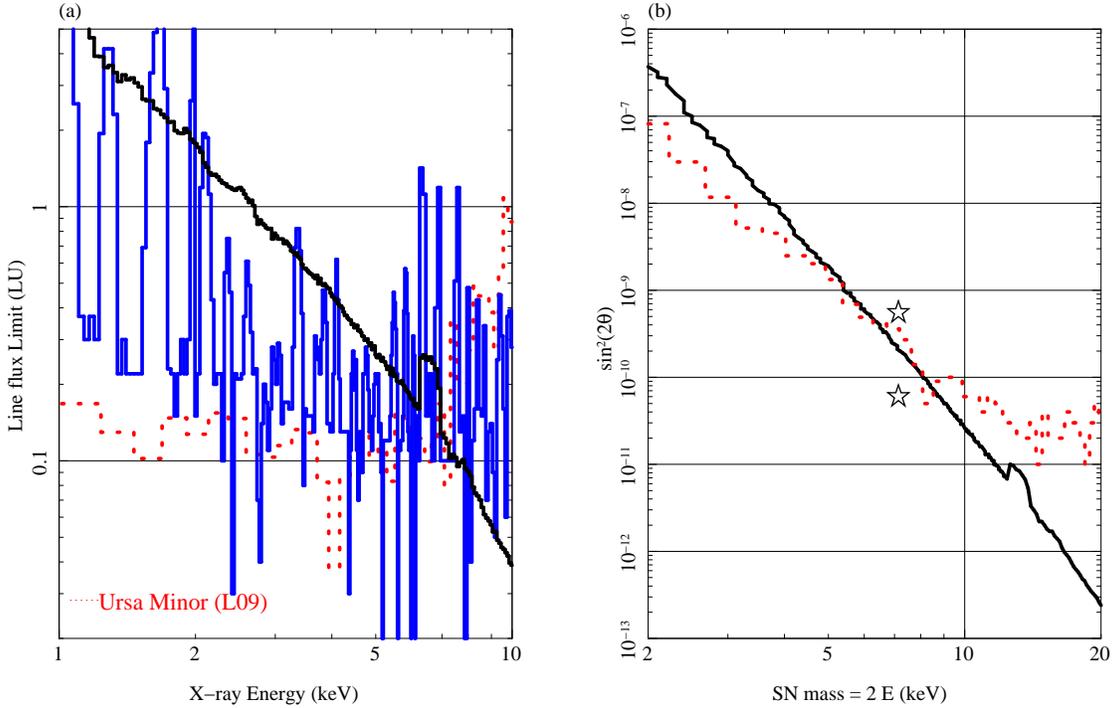

\begin{center}
\centerline{\hbox{
\includegraphics[scale=.38]{fig-energy-lu.ps}
\includegraphics[scale=.38]{fig-mass-sin3.ps}
}}
\caption{
(a) Limits on the line 
surface brightness 
in unit of photons cm$^{-2}$ s$^{-1}$ sr$^{-1}$ (line units, or LU).
The thick black line shows the XIS systematic upper limit
(1~eV in EW) from the Crab spectral fitting.
The blue histogram shows the 'observed' limits from the Perseus spectra.
The red dotted line shows the limit in L09 for the Ursa Minor dwarf galaxy.
(b) Limits on sterile neutrino mass and mixing angle.
Black and red dotted lines
show the XIS systematic upper limit
(1~eV in EW) 
and the limit in L09, respectively.
The 'detected' and constrained values in B14 are
indicated by star marks at 
mixing angles of
$55.3$ (Perseus, MOS), 
$6.0 $ (all other clusters, MOS)
all in $10^{-11}$ 
and at  mass of 7.1~keV.
}
\label{dis-dm:limit}
\end{center}
\end{figure}

In \S~\ref{crab-limit}, 
we evaluated the systematic sensitivity caused by instrumental calibration, mostly of the effective area,
based on the Crab Nebula spectral fitting.
We found the sensitivity to be 1--2~eV in EW over the 2.0--6.3~keV energy band.
Sensitivities below and above this band are expected to be larger than this level.
A constant systematic error of 1~eV EW is assumed and converted into line surface brightness 
in units of photons cm$^{-2}$ s$^{-1}$ sr$^{-1}$  (Line Unit =LU)
using the cluster continuum emission model, 
as presented in Fig.~\ref{dis-dm:limit}.
In \S~\ref{sect:fit-limit}, we constrained fluxes of non-detected line emission for a grid of energies
based on the deep XIS observations of the Perseus cluster center, 
using results not only in the 2.0--6.3~keV band but also in other bands (\S~\ref{sect:fit-lowhigh}).
We take the 1$\sigma$ upper limit from the Perseus C10 spectral fit 
(\S~\ref{sect:line-limit} and Fig.~\ref{fit-limit2:e-ew}).
When the upper limit is negative (i.e., absorption),
we approximate the line emission flux limit by '($1\sigma$ upper limit)-(best-fit flux)'.
Thus, obtained line flux values are converted into line surface brightnesses in the Fig.~\ref{dis-dm:limit}.
These 'observed' limits fluctuate largely on the energy
due to the photon statistics and any systematic effects.
In most energy ranges up to 5~keV
the systematic errors dominate the 'observed' limits.

We compare the current limit with that in \cite{Loewenstein2009} (L09)
who used the  \suzaku XIS observation of the Ursa Minor dwarf galaxy.
Contrary to the bright X-ray emission from the Perseus spectra,
the dwarf emission is negligibly faint.
This is a motivation for the target selection for L09.
In cases of faint X-ray sources such as the dwarf
systematic errors of effective area 
are less influential.
Therefore, 
their limit depends mostly on instrumental and cosmic background over the energy band
and is lower than our limit for $E<6$~keV.
Our deeper exposures minimize the difference in the statistical limit for $2<E<6$~keV.
At higher energies, both observations are limited by
the same instrumental background.
Based on similar motivations, \citet{Loewenstein2010} and \citet{Loewenstein2012}
used the ultra-faint dwarf galaxy Willman 1 with \chandra and \xmm, respectively.

We transform the line flux limit into 
the constraint on the mixing angle, $\sin^2 2\theta$, as a function of 
the dark matter particle mass, $m_{\rm SN}=2E$.
Here we take the cluster dark matter profile 
from B14 as
$1.49\times 10^{14}~M_\odot$ (projected mass) within a radius of 0.24~Mpc ($\sim 10'$)
and find a mass surface density ($\Sigma_{\rm DM}$) of 820 $M_\odot$ pc$^{-2}$.
We use the following relation
among line surface brightness ($f_{\rm SN}$), 
$\Sigma_{\rm DM}$, 
dark matter decay rate ($\Gamma$),
and the source redshift ($z$).

\begin{eqnarray}
f_{\rm SN} & = & \frac{\Sigma_{\rm DM} \Gamma}{4\pi (1+z)^3 m_{\rm SN}} \nonumber \\
           & \simeq & 
9.3\times 10^{-3} \frac{1}{(1+z)^3}
\left(\frac{\Sigma_{\rm DM}}{10^3 M_{\odot} {\rm pc}^{-2}}\right) 
\left(\frac{\Gamma }{10^{-30} {\rm s}^{-1}}\right)
\left(\frac{m_{\rm SN}}{{\rm keV}}\right)^{-1}
\, {\rm cm}^{-2} {\rm sr}^{-1} {\rm s} ^{-1} . \label{eq:sn_flux}
\end{eqnarray}
We take a decay rate $\Gamma$ from \cite{Loewenstein2010} as
\begin{eqnarray}
\Gamma =  1.38 \times 10^{-32} 
\left(\frac{\sin^2 2\theta}{10^{-10}}\right)
\left(\frac{m_{\rm SN}}{1 {\rm keV}}\right)^5 
\, {\rm s^{-1}}.
\label{eq:gamma-loe10} 
\end{eqnarray}

The result is shown in Fig.\ref{dis-dm:limit} as compared with that in L09.
Thanks to the larger dark matter mass in the Perseus cluster than that in the dwarf
the decay rate limits 
are comparable to or lower than those in L09
at least at $m_{\rm SN} >$ 4~keV.
We also indicate the 'detected' and constrained values from B14 (by star marks).
Consistently with the comparison in the line flux limits,
our decay rate limit at $m_{\rm SN}=7$~keV ($E=3.5$~keV)
is a few times smaller than that for the Perseus center with the MOS in B14.

There have been a number of X-ray measurements in the past on the sterile neutrino model.
Independently of these X-ray observations,
there are also other observational and theoretical constraints.
See the literature given in the Introduction 
for recent results, compilations, and detailed discussions.
Our limit at $4<m_{\rm SN}<$12~keV is comparable 
to the majority of previous X-ray limits.

\subsection{Prospects for the \astroh Dark Matter Search}

The joint JAXA/NASA project \astroh, to be launched in fiscal 2015, 
will be the first observatory providing high spectral resolution spectroscopy for extended sources
with an X-ray calorimeter (Soft X-ray Spectrometer; SXS).
See \citet{Takahashi2012} for the mission
and \citet{Mitsuda2010} for the SXS in detail.
B14 presented a 1~Ms SXS simulation of the Perseus cluster center
including their claimed line feature.
They found that the dark matter 3.5~keV line
can not only be detected 
but also 
distinguished from plasma features
by line broadening due to an assumed velocity dispersion of dark matter.

Previously in this section, we identified a number of factors limiting the line detection sensitivity.
For each factor, we now consider the \astroh aspect.
In \S~\ref{sect:dis} and \S~\ref{dis-cal}, 
instrumental features and calibrations are discussed.
Similarly to the \suzaku XIS, 
the \astroh SXS energy response
contains features such as O, Si and Au edges
and the effective area decreases towards lower and higher energies.
Thanks to the higher energy resolution,
effects of these instrumental features are localized into narrower energy channels (e.g. 10--20~eV).
The increased number of independent energy resolution elements
requires more detailed and better statistical calibration to maximize the SXS full capability. 

Because of line features in the cluster plasma and cosmic background
in the soft X-ray band we could search lines mainly at $E>2$~keV (\S~\ref{sect:dis}).
At $E<1$~keV, the complex cluster emission dominated by Fe-L transitions
combined with lower energy resolution (in terms of $E/\Delta E$) 
may not be fully resolved even with the SXS.
Conversely, 
above 1~keV the SXS will resolve many lines
not only from strong He-like triplets and Ly$\alpha$
but also from other weak transitions. 
These weak lines need to be modeled 
accurately in strength and position to separate them from any unidentified features.
These astronomical 'background' features can be avoided by using targets
with weak plasma emission.

Largely because of the small solid angle of SXS ($3'\times 3'$),
the SXS grasp and hence expected numbers of photon
are much smaller than for the XIS.
Thus, the photon statistics predominate in limiting the line sensitivity.
This disadvantage
can be partially compensated 
for by another \astroh instrument, the Soft X-ray Imager (SXI), 
with a large solid angle ($38'\times 38'$) and a moderate effective area.
Furthermore, the \astroh Hard X-ray Imager (HXI)
extends the line search above 10~keV up to 80~keV
with its spectral imaging capability.

The issue in modeling of adjoining lines and continuum emission discussed in \S~\ref{dis-com2}
will be removed by separating lines with the SXS.
For example,
the neighboring lines examined above have a separation larger than 100~eV in energy, 
which is more than 10 times larger than the SXS resolution.

In \citet{kitayama2014}, 
we studied and demonstrated the \astroh search for the dark matter signal.
We evaluated targets including X-ray bright clusters, 
the Milky Way Galactic center and bulge,
and dwarf galaxies.
Among them, clusters and the Galactic center will be primal targets for the \astroh early phase observations
for other purposes.
We will use these data to inspect claimed unidentified lines and search new features.
Once a hint of signal is detected, 
we can invest deeper exposures 
in the same targets or other X-ray faint objects.
The latter will provide higher dark matter signal-to-noise ratios.
Observations of objects at different redshifts or line widths (equivalently velocity dispersion)
are crucial for separating line origin from instrumental features.
Stacking analysis as performed in B14
will be more feasible with the improved energy resolution.
Finally, 
new types of targets
with massive dark matter but a shallower gravitational potential (for lower X-ray emission)
and small spatial extent, 
to be discovered from current and future galaxy surveys, 
together with the \astroh capabilities, 
will increase the possibility of discovering the dark matter signals.

\subsection{Summary}
\citet{bulbul2014} claimed detection of an unidentified line at 3.5~keV
not only from the Perseus center 
but also from other cluster data sets.
Using deeper \suzaku XIS data for the Perseus center,
we were unable to confirm this detection.
We have carefully examined systematic errors
from the effective area calibration using Crab Nebula observations.
The XIS upper limit of a line at the claimed position (3.51~keV)
is systematically determined at a half of the 'detected' best-fit value in B14, or 1.5~eV in EW.
This points to a similar level of systematic error in the analysis of B14.
In addition to possibilities argued in B14,
we suggest that the modeling uncertainty of the plasma continuum emission around the line
combined with instrumental calibration errors resulted in their detection.
We have also constrained unidentified line features over the 2.0--6.3~keV energy range.
These results from current CCD instruments 
indicate that distinguishing an unidentified line feature from any atomic lines from baryons 
or instrumental effects is challenging, 
largely because of limited energy resolution.
Higher resolution spectroscopy is the only way to resolve the unidentified line, 
separately from baryonic or instrumental effects,and confirm or reject its dark matter origin.

\bigskip
We thank the referee for useful suggestion
and the following scientists for helpful discussion,
N. Yoshida, A. Kamada, T. Kitayama,
K. Matsushita, E. Bulbul, R. Smith, A. Foster, 
M. Loewenstein, 
A. Boyarsky, 
O. Ruchayskiy, 
and N. Sekiya.
We thank all the \suzaku team member for their supports.
We acknowledge the support by a Grant-in-Aid for Scientific Research from the MEXT, No.24540243 (TT).

\appendix
\section{Fitting results of the Perseus C10 spectra in the low- and high-energy bands}
\label{sect:fit-lowhigh}

\begin{figure}[hpt]
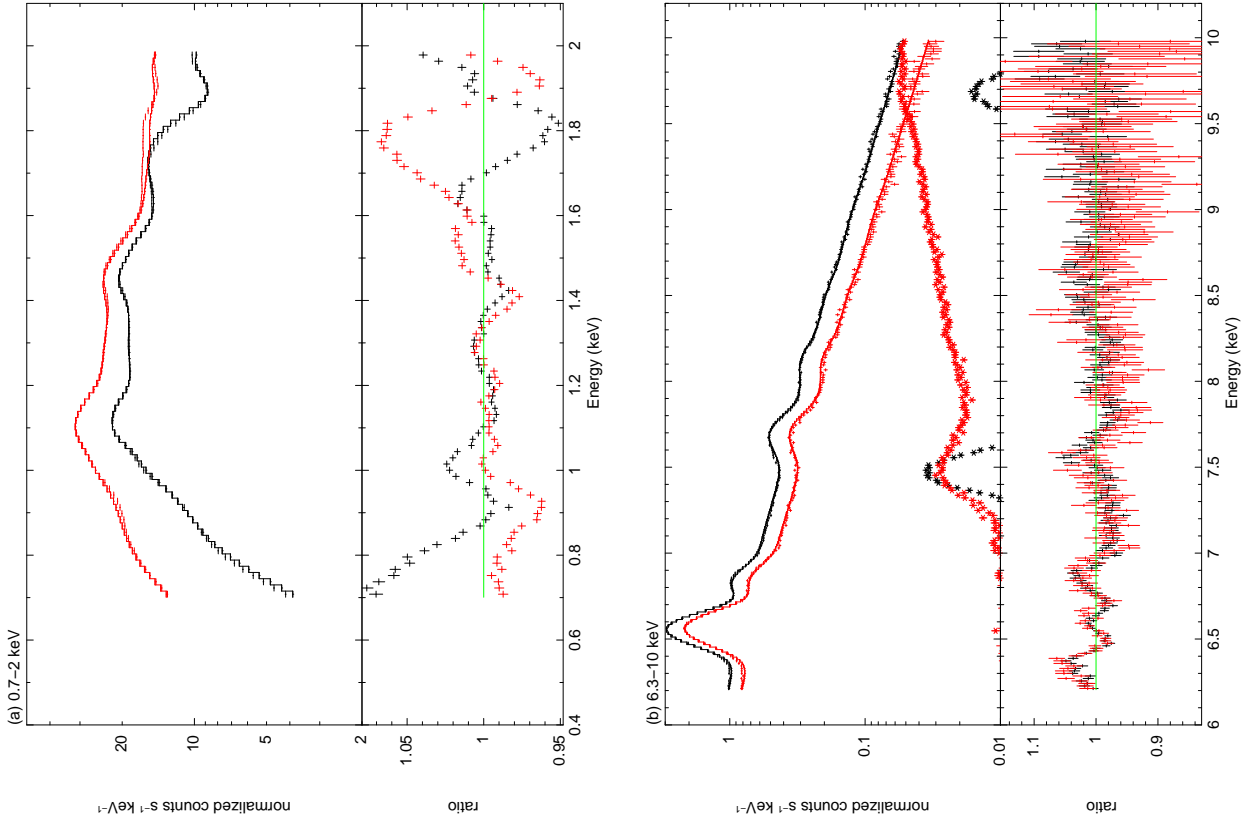

\begin{center}
\includegraphics[scale=.45]{fig-per-1008-1.ps}
\includegraphics[scale=.45]{fig-per-1009-1.ps}
\caption{
The XIS FI (black) and BI (red) spectra from the Perseus center (C10)
fitted with VAPEC models.
Panels (a) and (b) shows results in the low and high energy X-ray bands, 
respectively.
}
\label{fig:fit-lowhight}
\end{center}

\end{figure}

Fig.~\ref{fig:fit-lowhight}
shows fitting results in the low- and high-energy bands.
We use the VAPEC model along with 
Galactic photo-electric absorption (phabs model in XSPEC).

\end{document}